\documentstyle[aps]{revtex}

\begin{document}
\flushbottom

\def\thepage{\roman{page}}
\title{\vspace*{1.5in}COMBINED MULTI- FREQUENCY MAP FOR POINT SOURCE
SUBTRACTION
}

\author{P.Naselsky}

\address{Theoretical Astrophysics Center, Juliane Maries Vej
30, 2100 Copenhagen, {\O} Denmark\\ Rostov State University, Zorge 5, 344090
Rostov-Don, Russia\\ July 1997}

\author{D. Novikov}

\address{Astronomy Department, University of Oxford, NAPL, Keble
               Road, Oxford OX1 3RH, UK\\
Astro-Space Center of Lebedev Physical Institute, Profsoyuznaya 84/32,
Moscow, Russia\\
		email: novikov@astro.ox.ac.uk}

\author{Joseph Silk}

\address{Astronomy Department, University of Oxford, NAPL, Keble
               Road, Oxford OX1 3RH, UK\\
		email: silk@astro.ox.ac.uk}
\maketitle

\input{epsf}

\begin{abstract}
A method is proposed for combining  multi-frequency maps in order
to produce a catalogue of extragalactic point sources using 
data from future high-precision satellite experiments.
We have found  
the optimal way for combining maps at  different frequencies
in order
to maximize the  signal (point sources) to noise (rest of the signal) ratio.
Our approach is a natural  multi-frequency generalization of the band-pass filter
introduced by Tegmark and de Oliveira-Costa (1998).
We show that combination of different frequency maps gives us the possibility 
of creating a  more complete catalogue of point sources. 
\end{abstract}

\twocolumn
\section {Introduction}

In the next few years, the  new generation of CMB experiments (MAP and
PLANCK) will provide all-sky maps of the  cosmic microwave background (CMB) with
high resolutions and sensitivities.   
Theoretical investigations
of  different kinds of foregrounds, instrumental noise properties, scan strategies
and related topics for the future PLANCK mission are now in progress.  
The  PLANCK LFI and HFI instruments  will be able to measure the 
CMB anisotropy and polarization using 10 frequency channels that cover 
the frequency range 30 - 857 GHz (Mandolesi et al.,1998; Maino et al.,1999). 
The maps provided by PLANCK will contain contributions from various
physical components, including the primary and secondary CMB
signals and foregrounds such as free-free and synchrotron emission, 
the Sunyaev-Zeldovich 
effect, galactic dust and extragalactic point sources. One of the important
goals of this experiment, apart from the separation of the CMB from the
remaining parts of the signal, is the creation of a
point source catalogue.
The aim of this paper is to show how one may  optimize the strategy
for  producing  the point 
source catalogue by using the multi-frequency  properties of the radio/submillimetre sky
observations.      

The effect of point sources on satellite observations has been  widely discussed in  the
literature (see the  review by Vielva et al. (2001), and references therein). Many authors have developed various  linear and non-linear methods 
for point source
extraction. Hobson et al. (1999) suggested the combination of the  maximum-
entropy method (MEM) with  Mexican Hat wavelet filtering  in order to separate all the
physical components of the microwave ski including extragalactic point sources
and using different frequency channels. Tegmark and de Oliveira-Costa
(1998; hereafter TO-98)  described the linear filtering of point sources 
in the form of a band-pass filter in order to render  point sources on each
frequency map more visible relative to  the background of other physical components. 
Cayon (2000), 
Sanz et al. (2000) and Vielva et al. (2000) applied the Mexican Hat wavelet method
to  denoising  the CMB maps and for production 
of the Planck point source catalogue. Naselsky et al. (2000), Chiang et al.
(2001) proposed the approach of 
amplitude-phase analysis and applied it to  simulated  maps of the 
CMB  with  foregrounds, point sources and
pixel noise included  for the  accurate determination of 
the locations of bright point sources
and separation of these sources  from the rest of the signal by 
an iteration technique. 
All of  the methods mentioned above can complement each other
in the framework of the future highly sensitive CMB experiments because of the 
different sensitivities of  each  method  to different properties of the signal. 

In this paper we focus our attention on the combination of maps with different
frequencies in order to maximize the signal 
($S$- point sources) to noise ($N$- rest of the signal) ratio. 
We show that co-adding of harmonics from different maps with
specific weights
leads to an increase in   the
accuracy of the point source subtraction. The aim of our paper is
to produce the best possible combination of all frequency channels in order
to construct a map in which  the presence of point sources of
some particular population is most visible.
This procedure allows us to obtain a more complete and accurate catalogue
of point sources than can be  obtained by independently investigating   each frequency 
channel (TO-98).

The layout of our paper is as follows. In Section 2 we introduce the
definition of the 'best' map combination, derive
the  analytical solution for the maximum possible signal-to-noise ratio and
make some numerical predictions for the
Planck experiment. We also present
a possible special combinations of the multifrequency maps. 
We briefly summarize  our results in Section 3.

\section{OPTIMAL COMBINATION OF DIFFERENT FREQUENCY CHANNELS}
In this section, we derive the optimal way for  using the different
frequency channels to produce the best possible map for detection
and subsequent subtraction of extra-galactic point sources. The
idea behind  our approach is quite simple and transparent. We define the
'best' map as the combination of different frequency maps which maximizes the 
signal-to-noise ratio.
In our case, the signal is the signal from point sources. The noise
consists of other physical components including pixel noise,
CMB, dust, synchrotron and free-free emission. All of
these components
limit our capability to detect point sources. 

\begin{center}
{\it Optimization}
\end{center}

Suppose that we have $N$ different maps of the same region of the sky
made using $N$ different channels with  frequencies $\nu_{i},\hspace
{0.2cm} i=1,..,N$.
The flux of each point
source $S^j=S(\nu^j)$ is a function of frequency and, therefore,
differs from one
channel to another, while the position of this source is, obviously,
the same for all maps. In this subsection,  we  consider for simplicity one particular population of point sources with the same dependence of
fluxes on  frequency.
The contribution of point sources to the the temperature on the CMB map
in the  j-th frequency channel
can be written as a set of Dirac delta functions with coefficients
$S_i^j$ proportional to the flux of the source at  position 
$\overline{r}_i$:

\begin{equation}
\Delta T_{ps}^j=\sum\limits_{i=1}^{N_{ps}}S_i^j\delta(\overline{r_i},
\overline{r})
\end{equation}
Here, $N_{ps}$ is the total number of point sources.
The temperature fluctuations on the  map convolved with the antenna beam are
therefore: 

\begin{equation}
\begin{array} {c}
\Delta T^j(\overline{r})=\sum\limits_{i}s_i^jB^j(\overline{r_i}-\overline{r})+
\sum\limits_{lm}a_{lm}Y_{lm}(\overline{r})=\\
=\sum\limits_{lm}T_{lm}^jY_{lm}(\overline{r}),
\end{array}
\end{equation}
\begin{figure}[h]
\vspace{0cm}\hspace{0cm}\epsfxsize=9cm
\epsfbox{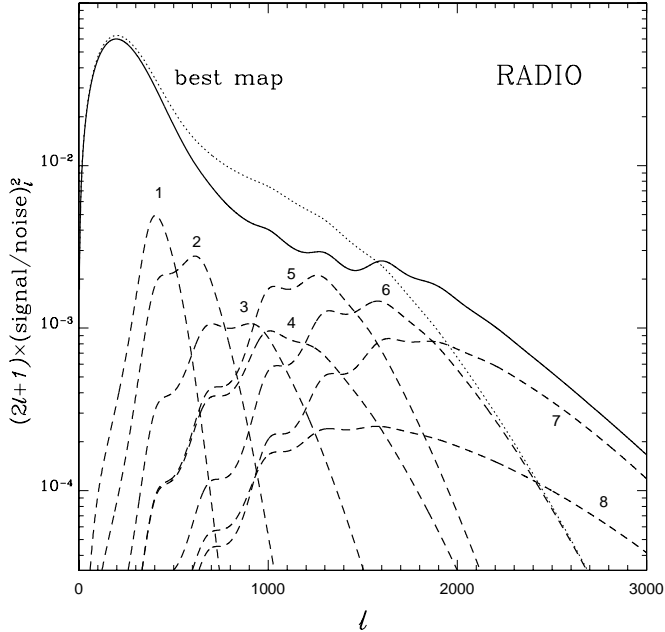}

   \caption{Signal-to-noise ratio (in arbitrary units) for each
channel shown separately after applying the TO-98 filter (dashed lines)
and for the combination of 10 Planck maps (solid line). The calculation
is performed for the population of radio sources with spectral
parameter $\alpha$ taken from Table 1. The dotted line shows the same ratio,
but for a break in the  spectrum between 143 and 217 GHz.
}
\end{figure}
where $B^j(\overline{r})$ is the beam for the j-th channel, 
the amplitude of the
point source $s_i^j$ is proportional to the flux 
and depends on the antenna width: $s_i^j\sim S_i^j/\Theta^2$. 
The coefficients
$a_{lm}^j$ have zero mean and variance
$\langle (a_{lm}^j)^2\rangle =C_{lm}^j(b_{lm}^j)^2+C_{pix}^j$ 
(pixel noise is not
convolved with the beam). The set of harmonics $T_{lm}^j,\hspace{0.5cm}j=1,..,N$
can be considered as the N-dimensional vector: $\overline{T}_{lm}$. The
aim of our paper is to combine $N$ given maps with known values
of the expected noise to construct the map with the best signal-to-noise ratio. In order to produce such a map, we 
introduce the vector-filter $\overline{f}_{lm}$ with components 
$f_{lm}^1,..,f_{lm}^N$ and consider the combination of N maps in the
following form:

\begin{equation}
\begin{array} {l}
\Delta \widetilde{T}=\sum\limits_{lm}\overline{f}_{lm}^{T}\overline{T}_{lm}
Y_{lm}(\overline{r})
=\sum\limits_{lm}\widetilde{T}_{lm}Y_{lm}(\overline{r})
\end{array}
\end{equation}
where $\widetilde{T}_{lm}=\sum\limits_{j=1}^{N}f_{lm}^jT_{lm}^j$. 
We consider a symmetric antenna beam and subsequently drop  the subscript m because
$\langle\left( \overline{T}_{lm}\right)^2\rangle=
 \langle\left( \overline{T}_{l}\right)^2\rangle$. Thus 
the filter that we use
should depend on l only: $\overline{f}_l=\overline{f}_{lm}$. Our approach can readily be
generalized to the case of an anisotropic beam, but this is beyond the scope of
the present  paper.
Our problem now is to find vectors $\overline{f}_l$ for each $l$ so that the 
ratio $\gamma=signal/noise$ is maximized:
\begin{figure}[h]
\vspace{0cm}\hspace{0cm}\epsfxsize=9cm
\epsfbox{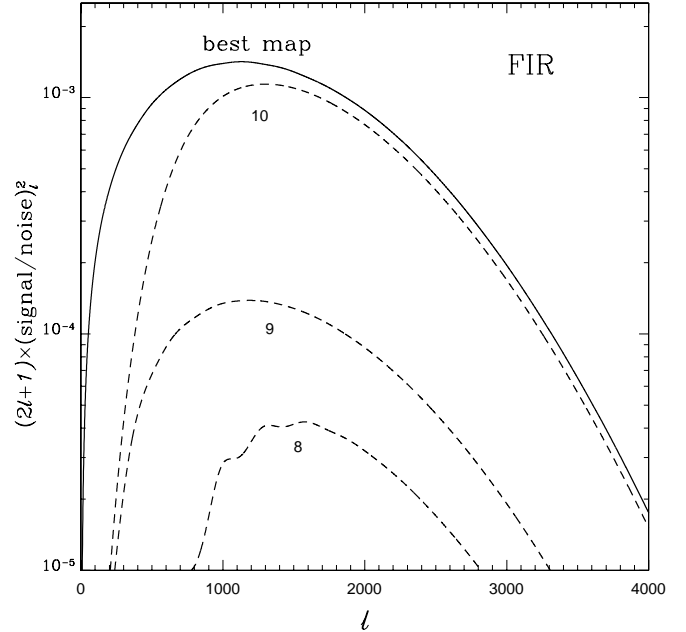}
    \caption{Same as fig.1 but for the population of FIR sources}
\end{figure}

\begin{equation}
\begin{array} {c}
\gamma=\Delta \widetilde{T}_{ps}(\overline{r}_i)/\widetilde{\sigma}_{tot},\\
\Delta \widetilde{T}_{ps}(\overline{r}_i)=\sum\limits_{l}
(2l+1)\overline{\beta}_l^{T}\overline{f}_l,\\
\widetilde{\sigma}_{tot}^2=\langle\Delta \widetilde{T}_{tot}^2\rangle=
\sum\limits_{l}(2l+1)\overline{f}_l^{T}M_l\overline{f}_l
\end{array}
\end{equation}

Here $\overline{\beta}_l$ is the vector with components
$s_i^jb^j,\hspace{0.5cm}j=1,..N$ 
($b_l$ is the beam) and $M_l=[m_l^{j_1,j_2}]$ 
is the covariance matrix:
$m_l^{j_1,j_2}=\langle a_{lm}^{j_1}a_{lm}^{j_2}\rangle$. 
In our
calculations, we use:

\begin{equation}
m_l^{j_1,j_2}=\left(C_l^{j_1}C_l^{j_2}\right)^{1/2}b_l^{j_1}b_l^{j_2}+
C_{pix}^{j_1}\delta_{j_1}^{j_2}
\end{equation}
where $\delta_{j_1}^{j_2}$ is the Kronecker symbol.
The ratio $\gamma$ from equation (4) is maximized if we choose
the filter to be:

\begin{equation}
\overline{f}_l=M_l^{-1}\overline{\beta}_l
\end{equation}
Consequently, one can
conclude that point
sources with some particular dependency of  flux on frequency
 are most clearly seen in the map that
is the combination of $N$ maps taken with these specific weights of 
harmonics.
If only one channel is under consideration, then this
formula becomes especially simple and is exactly the same as
the one obtained by Tegmark and de Oliveira-Costa (TO-98):

\begin{equation}
f_l\sim\frac{b_l}{b_l^2C_l+C_{pix}}
\end{equation}

Using equations (4) and (6), one can get the final result for
the maximum possible signal-to-noise ratio $\gamma$, which is
achieved in the combined map:

\begin{figure}[h]
\vspace{-3cm}\hspace{0cm}\epsfxsize=9cm
\epsfbox{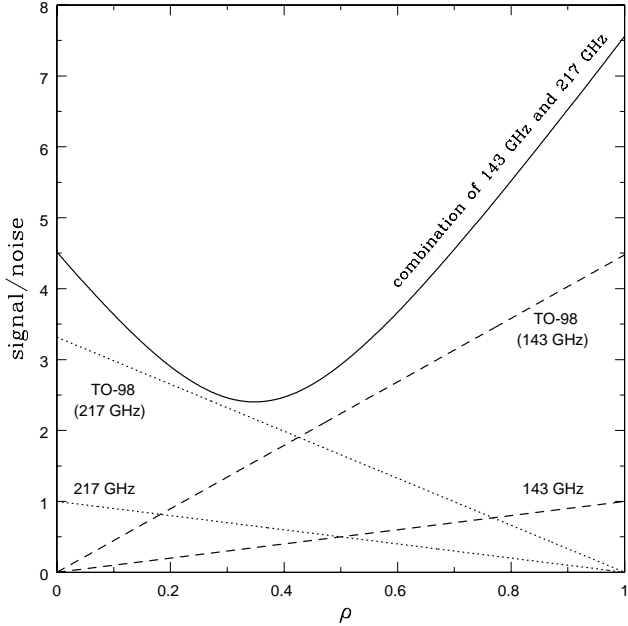}
    \caption{Signal-to-noise ratio for 143 GHz and 217 GHz channels 
after performing TO-98 filtering and for the combination of channels.
Lower straight lines show signal/noise for initial maps (in units
of $\sigma_{tot}$) while
upper straight lines show this ratio after TO-98 filtering.
Dotted lines: 217 GHz, dashed lines: 143 GHz,
solid line: combination of maps.}
\end{figure}

\begin{equation}
\gamma=\left(\sum\limits_{l}(2l+1)\overline{\beta}_l^{T}
M_l^{-1}\overline{\beta}_l\right)^{1/2}
\end{equation}

\begin{center}
{\it Estimates for the Planck experiment}
\end{center}

A multi-frequency analysis gives us the  possibility of detecting point
sources in coincidence with different frequency maps. 
In our calculations, we assume the frequency dependence of the intensity for
some particular point source population in the form of a simple
power law: $\nu^\alpha$. Multiplying this intensity by the conversion
factor between surface brightness and temperature 
(Tegmark and Efstathiou, 1996) and taking into account the resolution
$\Theta$, one can obtain  the dependence of the point source temperature 
on  frequency: 

\begin{equation}
\Delta T_{ps}^j\sim\nu_j^\alpha\frac{sinh(\nu_j/56.8GHz)}{\Theta_j^2\nu_j^4},
\hspace{1cm}j=1,..,10.
\end{equation}
where $j$ denotes the Planck channel number. Channels are numbered in
order of increasing frequency (see Table 1,2). Two different populations
have been chosen: radio and far-IR sources. Up to about 200 GHz, we can
expect mostly radio sources to be detected, while at higher frequenies
far-IR sources dominate. For the radio source  population, we took the spectral
index $\alpha$ from (Vielva et al., 2001) to be close to zero up to
100-200 GHz (table 1) and increasing to $-\infty$ at higher frequency.
For the far-IR population, this spectral index has the opposite behavior
and we choose $\alpha$  to be $\approx 2.7$ for 857-352 GHz  with practically
no contribution from these sources to the signal at lower frequencies.   
Of course, our estimates are   applicable only for these two specific
populations. Each particular population of point sources
should be investigated separately. This means that we should 
perform  the combination of channels with specific filters $\overline{f}_l$ which
correspond to the appropriate spectral index $\alpha$ for this population.

Using the results of the previous subsection, one can calculate 
the signal-to-noise ratio for each channel separately 
$\gamma(j)$ (if only the j-th channel is under consideration (TO-98)) and for
the combination of N channels. From eq. [7,8] one finds:

\begin{equation}
\begin{array} {c}
\gamma(j)_l^2=(\beta_l^j)^2/m_l^{j,j},\\
\gamma_l^2=\overline{\beta}_l^{T}M_l^{-1}\overline{\beta}_l,\\
\gamma^2=\sum\limits_{l}(2l+1)\gamma_l^2.
\end{array}
\end{equation}
The contribution to $signal/noise$ in each $l$ is obviously higher
for the combination of maps than for each map separately and
consequently $\gamma>\gamma(j)$. 
In fig. (1,2) we show this ratio for two populations (radio and FIR)
as described above. A significant improvement can be achieved by
 considering both  populations, $\gamma=\lambda\times max[\gamma (j)]$,
j=1,..,10. For radio source,s we found $\lambda\approx 4.4$
and for FIR $\lambda$ is about 1.2. Therefore, a more complete catalogue
 can be produced for this population of sources. 

\begin{center}
{\it Special cases and a possible combinations }
\end{center}

Let us for simplicity consider a special case by dealing
with only two channels with different frequencies: $\nu_1,\nu_2$.
Since the spectral behavior of different populations of point
sources is not well known, we introduce the parameter $\rho$. This parameter
contains  information about amplitudes of temperature fluctuations
due to the same point sources in   different channels:

\begin{equation}
\Delta T_{ps}^1=\rho\sigma_{tot}^1,\hspace{0.8cm}
\Delta T_{ps}^2=(1-\rho)\sigma_{tot}^2
\end{equation}
In fig. 3 we show the signal-to-noise ratio that can be found
applying TO-98 filters for the 143 and 217 GHz channels separately
and for the combination of these two maps with different possible
ratios of amplitudes $\Delta T_{ps}^1$ and $\Delta T_{ps}^2$. We choose
these two channels because significant steepening
 or even a break in the point source spectrum due to synchrotron emission is expected over  this
range of frequencies, so   it is hard to predict the  exact value of $\rho$.

For smaller values of $\nu$ quite a large number of point sources are
expected to be found in coincidence for different channels. It is interesting
to investigate how does our filter look like if we combine for example
the 70 GHz and 100 GHz (LFI) maps.  
\begin{figure}[h]
\vspace{-1cm}\hspace{0cm}\epsfxsize=9cm
\epsfbox{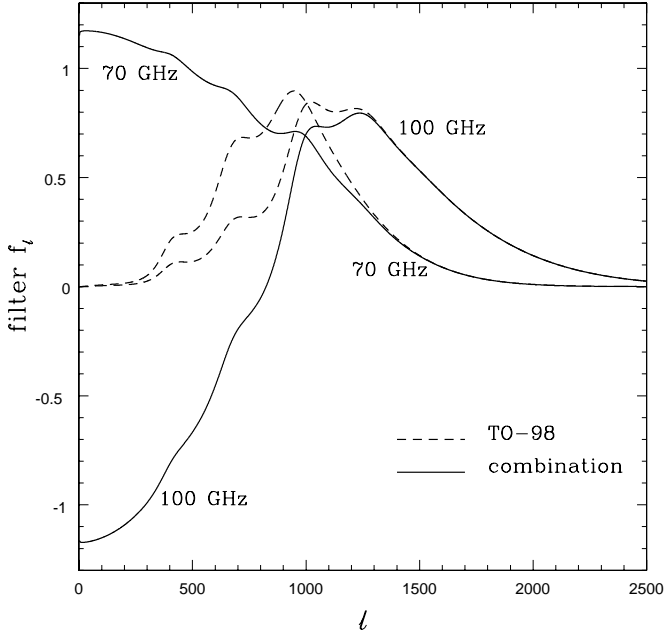}
    \caption{ Components of the filter $\overline{f_l}$ 
 for  combining the best possible map from 
the 70 and 100 (LFI) GHz channels
 (solid lines) in the comparison with TO-98 filters for each map 
taken separately
(dashed lines). The filter is given in arbitrary units.}
\end{figure}

According to formula (6) one can obtain the filter $\overline{f_l}$
with components $f_l^1,f_l^2$: 

\begin{equation}
f_l^1=\frac{\beta_l^1m_l^{22}-\beta_l^2m_l^{12}}{det(M_l)},\hspace{1cm}
f_l^2=\frac{\beta_l^2m_l^{11}-\beta_l^1m_l^{12}}{det(M_l)}
\end{equation}
where $det(M_l)=m_l^{11}m_l^{22}-(m_l^{12})^2$. Using eq [5] we
can investigate the behaviour of these two components at different scales. 

For small $l$'s
that correspond to  scales much larger than the antenna resolution,
the instrumental noise is negligible (at least for the
low frequency channels:
$\nu<353GHz$) and:

\begin{equation}
\begin{array} {c}
\vspace{0.5cm}
m_l^{11}\approx C_l^1(b_l^1)^2\approx C_l^1,\hspace{1cm}
m_l^{22}\approx C_l^2(b_l^2)^2\approx C_l^2,\\
m_l^{12}=(C_l^1C_l^2)^{1/2}b_l^1b_l^2\approx (C_l^1C_l^2)^{1/2}
\end{array}
\end{equation}
Substituting [13] in [12] we obtain:

\begin{equation}
\begin{array} {c}
f_l^1=\frac{\sqrt{C_l^1}}{det(M)}\left(\beta_l^1\sqrt{C_l^1}-
\beta_l^2\sqrt{C_l^2}\right),\\
f_l^2=\frac{\sqrt{C_l^2}}{det(M)}\left(\beta_l^2\sqrt{C_l^2}-
\beta_l^1\sqrt{C_l^1}\right).
\end{array}
\end{equation}
Therefore, harmonics with small $l$ from different channels are taken with
the opposite sign. Thus the contributions from the first and second maps to the
combined map compensate each other. Finally the  total signal in the
combined map turns to be close to zero at large scales. 
This means that our filter removes contributions from the CMB 
and other physical components at scales larger that the typical scale
of the point sources.

For $l$ values that  correspond to  scales comparable with
the  antenna beam,
the pixel noise is dominant and:

\begin{equation}
m_l^{11}\approx C_{pix}^1,\hspace{0.8cm}
m_l^{22}\approx C_{pix}^2,\hspace{0.8cm}
m_l^{12}\approx 0
\end{equation}
Therefore the components of the filter $\overline{f}_l$ are close to
the filters found in TO-98 for each channel separately (see fig.4):

\begin{equation}
f_l^1\approx \frac{\beta_l^1}{C_{pix}^1},\hspace{0.8cm}
f_l^2\approx \frac{\beta_l^2}{C_{pix}^2}
\end{equation}
These components are both positive and add to  form  peaks that correspond
to point sources in the combined map.

\section{Conclusions.}

We have presented an analysis of microwave sky observations based on the combination
of channels with different frequencies 
in order to separate extragalactic
point sources for subsequent production of a point source  catalogue.  
We made some numerical estimates for the future Planck
experiment and showed that a multi-frequency analysis allows us
to increase the signal-to-noise ratio by a factor of 1.2-4.4 to create
a more complete and precise catalog of point sources. 
{\onecolumn
\begin{table}
\begin{center}
\begin{tabular}{cccccc}
Channel   &Frequency& spectral index &signal/noise &signal/noise& signal/noise\\
number (j)& (GHz)   &   ($\alpha$)   &  (initial)  & (TO-98)    &(combination)\\
\hline
1         &  30     &    0.08        &    1.000      & 3.083    &  3.066 
 \\
2         &  44     &    0.08        &    0.986      & 3.091    &  5.894
 \\
3         &  70     &   -0.10        &    0.935      & 2.495    &  8.579
 \\
4         &100(LFI) &   -0.16        &    0.947      & 2.490    &  10.862
 \\
5         &100(HFI) &    -----       &    0.872      & 3.622    &  13.026
 \\
6         & 143     &   -0.55        &    0.766      & 3.429    &  14.858
 \\
7         & 217     &   -0.55        &    0.952      & 3.157    &  15.747
 \\
8         & 353     &   -0.55        &    0.895      & 1.886    &  15.834
 \\
9         & 545     &   $-\infty$    &    0.000      & 0.000    &  15.847
 \\
10        & 857     &    -----       &    0.000      & 0.000    &  15.925
 \\
\end{tabular}
\end{center}
\caption{Results of combination of  10 Planck channels for the population of
radio sources with given spectral behavior. The frequency of each channel 
is indicated in column 2 for both LFI and
HFI. The 3-rd column gives spectral indices as calculated by Vielva et al. (2001). 
$-\infty$ denotes spectral break. Fourth column shows the initial signal-to-noise 
ratio in units of $\sigma_{tot}^j$ under the condition that the
point source in the 1-st channel has $1\sigma_{tot}^1$ level.
The fifth column gives this ratio after applying the TO-98 filter to each channel
separately. In the last column we show
this ratio for the best combination of channels from 1 to j. This means that
the 2-nd row gives the  result for the combination of the 1-st and 2-nd channels,
3-d means the combination of 1,2,3 channels and so on.}
\end{table}}
{
\begin{table}
\begin{center}
\begin{tabular}{cccccc}
Channel   &Frequency& spectral index &signal/noise &signal/noise& signal/noise\\
number (j)& (GHz)   &   ($\alpha$)   &  (initial)  & (TO-98)    &(combination)\\
\hline
10        & 857     &    2.46     &    1.000      & 6.281    &  6.281 
 \\
9         & 545     &    2.71     &    0.811      & 2.257    &  6.575
 \\
8         & 353     &    2.71     &    0.551      & 1.162    &  6.960
 \\
7         & 217     &   $\infty$  &    0.000      & 0.000    &  7.405
 \\
6         & 143     &    -----    &    0.000      & 0.000    &  7.460
 \\
5         & 100(HFI)&    -----    &    0.000      & 0.000    &  7.468
 \\
4         & 100(LFI)&    -----    &    0.000      & 0.000    &  7.470
 \\
3         & 70      &    -----    &    0.000      & 0.000    &  7.470
 \\
2         & 44      &    -----    &    0.000      & 0.000    &  7.470
 \\
1         & 30      &    -----    &    0.000      & 0.000    &  7.470
 \\
\end{tabular}
\end{center}
\caption{Same as Table 1 but for FIR sources population. The channels are shown
in the opposite order. Initial signal/noise ratios are given under the assumption that the 
amplitude of the source in  the 10-th channel is $1\sigma_{tot}^{10}.$}
\end{table}}

\begin{center}
{\bf Acknowledgments}
\end{center}
 
This investigation was partly supported by INTAS under grant number
97-1192, by RFFI under grant 17625 and by Danmarks Grundforkskningfond
through its support for TAC.

\end{document}